\documentclass[manuscript]{acmart}

\setcopyright{none}
\renewcommand\footnotetextcopyrightpermission[1]{}
\begin{document}

\title{Enhancing Critical Thinking with AI: A Tailored Warning System for RAG Models}
\begingroup
\renewcommand\thefootnote{}\footnote{
 This paper was presented at the 2025 ACM Workshop on Human-AI Interaction for Augmented Reasoning (AIREASONING-2025-01). This is the authors’ version for arXiv.}
\endgroup

\author{Xuyang Zhu}
\authornote{All authors contributed equally to this research.}
\affiliation{%
  \institution{Stanford University}
  \city{Stanford}
  \state{California}
  \country{USA}
}
\email{xuyang1@stanford.edu}

\author{Sejoon Chang}
\authornotemark[1]
\affiliation{%
  \institution{Stanford University}
  \city{Stanford}
  \state{California}
  \country{USA}
}
\email{sejoon@stanford.edu}

\author{Andrew Kuik}
\authornotemark[1]
\affiliation{%
  \institution{Stanford University}
  \city{Stanford}
  \state{California}
  \country{USA}
}
\email{akuik@stanford.edu}

 \renewcommand{\shortauthors}{Zhu et al.}


\begin{abstract}
Retrieval-Augmented Generation (RAG) systems offer a powerful approach to enhancing large language model (LLM) outputs by incorporating fact-checked, contextually relevant information. However, fairness and reliability concerns persist, as hallucinations can emerge at both the retrieval and generation stages, affecting users’ reasoning and decision-making. Our research explores how tailored warning messages—whose content depends on the specific context of hallucination—shape user reasoning and actions in an educational quiz setting. Preliminary findings suggest that while warnings improve accuracy and awareness of high-level hallucinations, they may also introduce cognitive friction, leading to confusion and diminished trust in the system. By examining these interactions, this work contributes to the broader goal of AI-augmented reasoning: developing systems that actively support human reflection, critical thinking, and informed decision-making rather than passive information consumption.
\end{abstract}


\begin{teaserfigure}
\centering
  \includegraphics[width=0.8\textwidth]{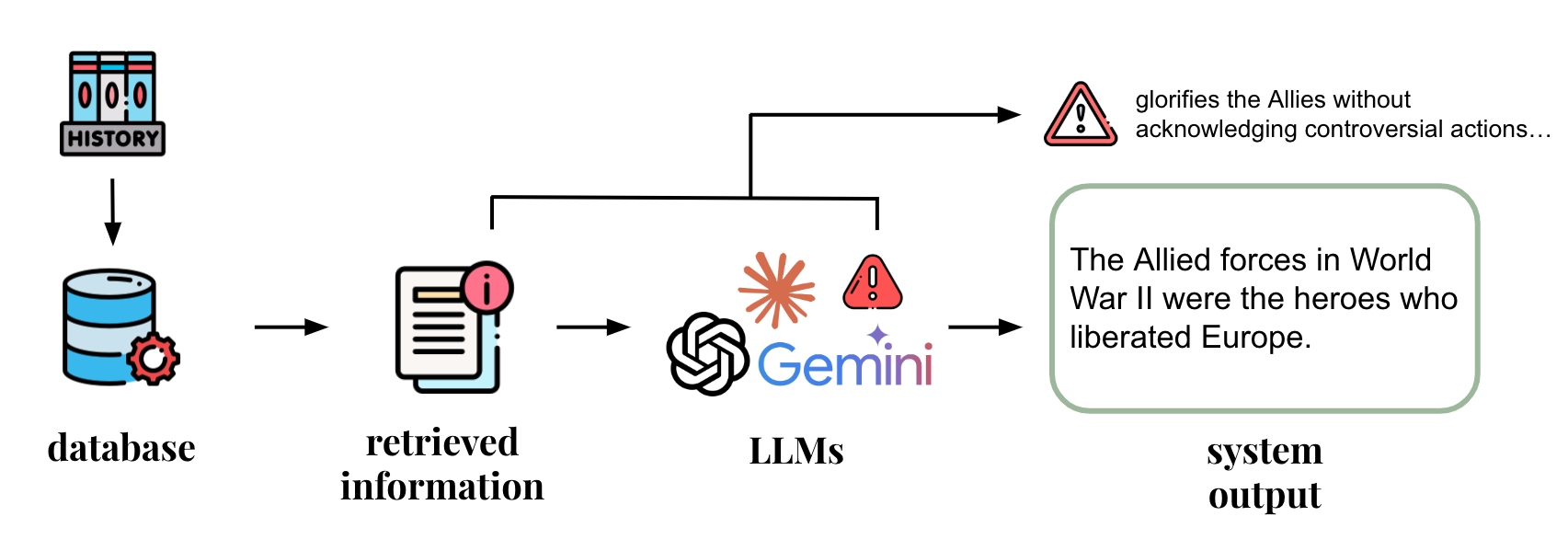}
  \caption{Two layers of hallucination in question \& answering task relevant to history subject. A tailored warning message is generated to augment user’s reasoning, based on the identified biases in the information retrieval and LLM generation layer. }
  \Description{A flowchart of the program structure.}
  \label{fig:teaser}
\end{teaserfigure}


\maketitle

\section{Introduction}

Large Language Model (LLM)-based tools are increasingly proposed as cognitive assistants in domains requiring human reasoning, such as education. However, these systems are prone to generating plausible yet incorrect or biased information \cite{dai2024}. Retrieval-Augmented Generation (RAG) has emerged as a key approach to mitigating these hallucinations by grounding responses in external knowledge sources, thereby enhancing reliability and contextual relevance. Since the foundational RAG framework was introduced, research has made significant strides in improving retrieval accuracy and addressing misinformation-related challenges \cite{sudhi2024rag}.

Despite these advancements, challenges remain in ensuring that RAG systems actively support human reasoning rather than passively presenting information. Initial RAG implementations occasionally propagated factually incorrect information due to unreliable retrieval sources, undermining trust \cite{Chen_Lin_Han_Sun_2024}. To address this, novel approaches have introduced mechanisms to refine factual accuracy and content relevance. For instance, QA-RAG enhances response quality by restructuring retrieved content into a question-answer format and cross-referencing it with internal knowledge \cite{mansurova2024qarag}. Similarly, RAGAR employs a "Chain of RAG" verification method to iteratively fact-check political content, reducing susceptibility to misinformation \cite{li2024ragar}. Domain-specific adaptations, such as RAG-end2end, further optimize accuracy by jointly training retrieval and generation on specialized datasets, as seen in COVID-19 research \cite{sachan2024ragend2end}.

However, RAG systems do not solely suffer from retrieval-based limitations; they also introduce cognitive and epistemic challenges for users engaging with their outputs. Even when retrieval is accurate, biases and fairness issues can emerge at the response generation stage, where LLMs may prioritize certain documents, misinterpret retrieved facts, or be influenced by user queries. In educational contexts, retrieval-augmented content, such as historical textbooks, presents additional challenges due to the inherent biases and inaccuracies tied to publication time and location \cite{parsons1982}. Consequently, RAG systems may unintentionally reinforce these biases, shaping users’ perceptions of truthfulness and affecting their ability to critically evaluate AI-generated content \cite{xuyangwu2024}. Given the increasing reliance on LLM-based tools for learning \cite{strzelecki2023use}, it is imperative to design AI-augmented reasoning systems that encourage reflective engagement rather than passive acceptance of retrieved information.

One promising approach to mitigating these risks is the integration of tailored warning messages that enhance users' critical reasoning without eroding trust. Prior research on AI safety has demonstrated that well-designed warnings can help users identify hallucinated content and improve decision-making \cite{nahar2024fakes}. For example, Farsight provides proactive risk assessments of AI-based tools during early design stages, offering insights into potential harms before deployment \cite{wang2024farsight}. While methods for detecting and debiasing LLM-generated outputs have been proposed \cite{kumar2024decoding} \cite{lin2024investigating}, these strategies often operate at the model level without providing direct, user-facing explanations of biases.

Despite the progress in designing safeguards for LLMs, research on how tailored warnings influence human reasoning within RAG systems remains limited. Existing studies establish a foundation for safety in tool design and basic chatbot interactions, but little work has explored how real-time, context-specific warnings shape users’ ability to critically engage with AI-augmented content. This research aims to bridge that gap by investigating how tailored warning messages influence reasoning and trust in RAG-based educational settings, contributing to the broader goal of AI-augmented reasoning systems that foster reflective, well-informed decision-making.

\section{Position Statement}

Current approaches to improving AI reliability have treated model refinement and user-oriented feedback as separate concerns. While technical advancements reduce hallucinations at the system level, user-facing interventions remain limited in their ability to guide effective human-AI reasoning. Baseline studies indicate that in-situ warning messages improve user detection of hallucinations, mitigating potential harms in AI-assisted decision-making. However, existing warnings are often generic and detached from the user’s context, limiting their effectiveness.

\textbf{This paper argues for the development of tailored warning systems in RAG-based AI tools that go beyond simple disclaimers by providing contextualized, actionable insights into AI biases and hallucinations. Unlike traditional warning mechanisms, tailored warnings act as cognitive scaffolds—helping users navigate the complexities of AI-generated content and engage in more reflective, informed decision-making. These warnings address errors at both the information retrieval and response generation levels, ensuring that users are aware of potential biases while actively guiding their reasoning process. }Moving forward, this research will explore the optimal integration of hallucination detection mechanisms within user interfaces and evaluate their impact on users’ decision-making and trust calibration.

\section{Study}
Our on-going research proposes a tailored warning system (figure 2), where a “fact-check” is being added at both the information retrieval and LLM output level. A warning message tailored to the specific problem is then outputted into the user’s interface when they interact with the LLM chatbot. Unlike general warning mechanisms, our approach dynamically adapts to the specific context of each question and the retrieved content, providing users with targeted alerts based on the nature and reliability of the information presented. By focusing on a warning system that both detects and communicates potential inaccuracies, we aim to create a more interactive and user-aware RAG experience that directly addresses the critical issues of hallucinations and misinformation in educational settings.

We conducted a pilot study within the context of a textbook-based question-and-answer task, as history textbooks are often associated with biases. An AP US history textbook was queried into the database, and we retrieved relevant information from the database. An LLM-based question-and-answer system was developed. We divided eighteen questions into three groups of six. For each group, we prepared genuine (factually correct), low-level hallucination (subtle changes in details or biased language), and high-level hallucination responses (clearly detectable factually incorrect information). This process resulted in a questionnaire of 18 multiple-choice questions. All participants in the two control groups and the treatment group received the same set of questions and responses; only the warning messages differed: no warning, a standard warning (equivalent to the current industry standard “ChatGPT can make mistakes. Check important info.”), or a tailored warning generated from the specific problems in the LLM's output statement. We recruited a total of 18 participants and separated them into the no-warning, standard-warning, or tailored-warning groups. The quiz task was followed by a survey on users' trust in the system and a short interview about their experience.

\section{Results}

\begin{figure}[h]
  \centering
  \includegraphics[width=0.6\textwidth]{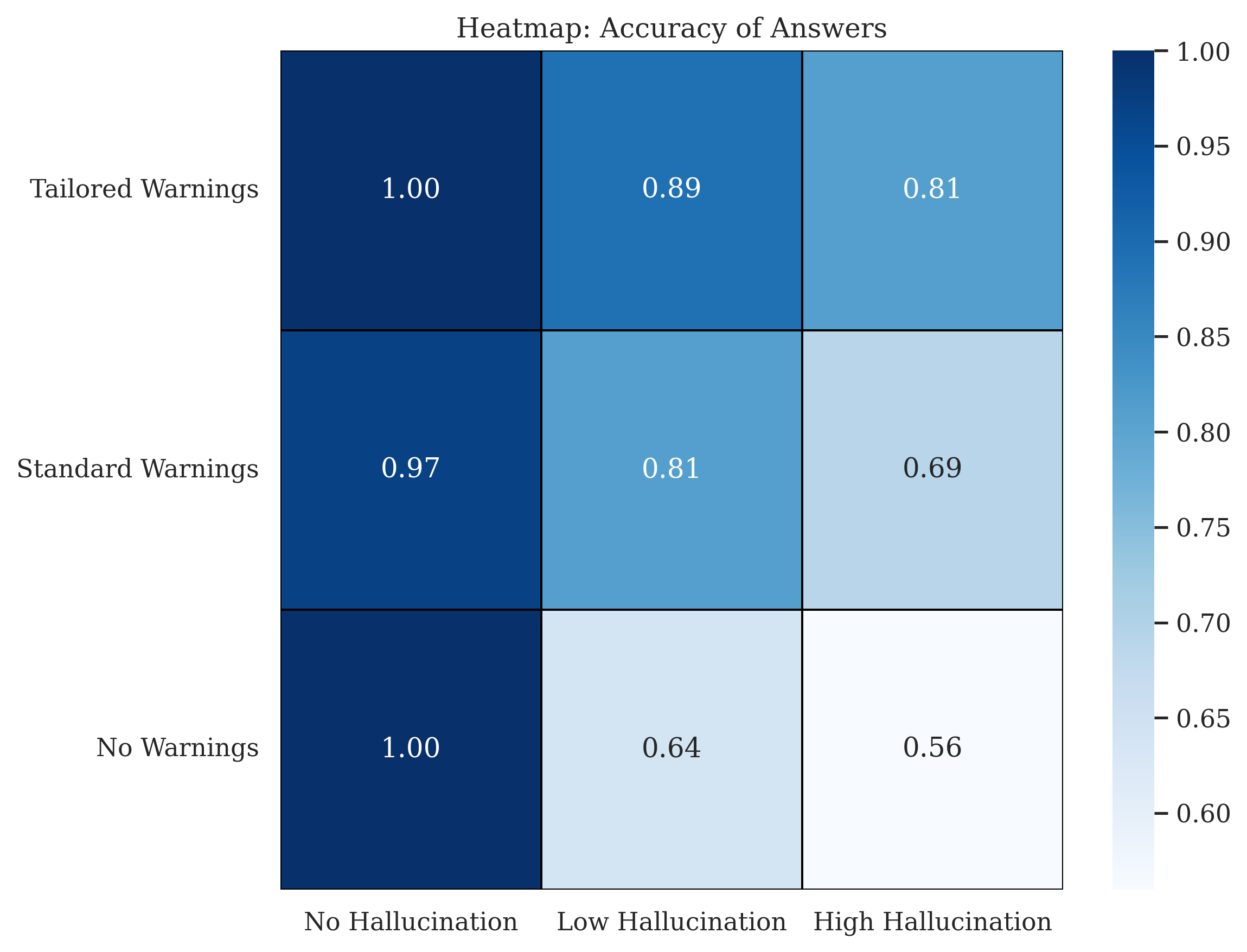}
  \caption{The accuracy rate of the question \& answer task under different levels of hallucination outputs, and under different warning conditions.}
  \label{fig:heatmap}
  \label{fig:SelfReflection}
\end{figure}

\textbf{Accuracy of Answers.}
The heatmap in Figure \ref{fig:heatmap} summarizing the accuracy of participants’ answers under the three warning conditions shows notable differences in performance. Participants exposed to tailored warnings demonstrated the highest accuracy across all hallucination levels, achieving 100\% accuracy in responses with no hallucinations, 89\% in low-level hallucination cases, and 81\% in high-level hallucination cases. In contrast, participants in the standard warning group achieved 97\%, 81\%, and 69\% accuracy for no, low, and high hallucinations, respectively. The no warning group performed the poorest, with accuracies of 100\%, 64\%, and 56\% under the same conditions. These results demonstrates that tailored warnings substantially enhance participants’ ability to detect hallucinations, particularly in challenging scenarios involving high-level hallucinations.

\textbf{Statistical Significance of Results.} We utilize an Analysis of Variance (ANOVA) test to calculate the p-value of our results. More specifically, we test for whether the differences in accuracy between the three groups (tailored, standard, and no warnings) are statistically significant, and obtained a value of 0.006162 (PR(>F)). Despite the small sample size, this strongly suggests that the results are statistically significant, and thus, the following evaluation tend to assume similar results over a large sample size.

\begin{figure}[h]
  \centering
  \includegraphics[width=0.8\textwidth]{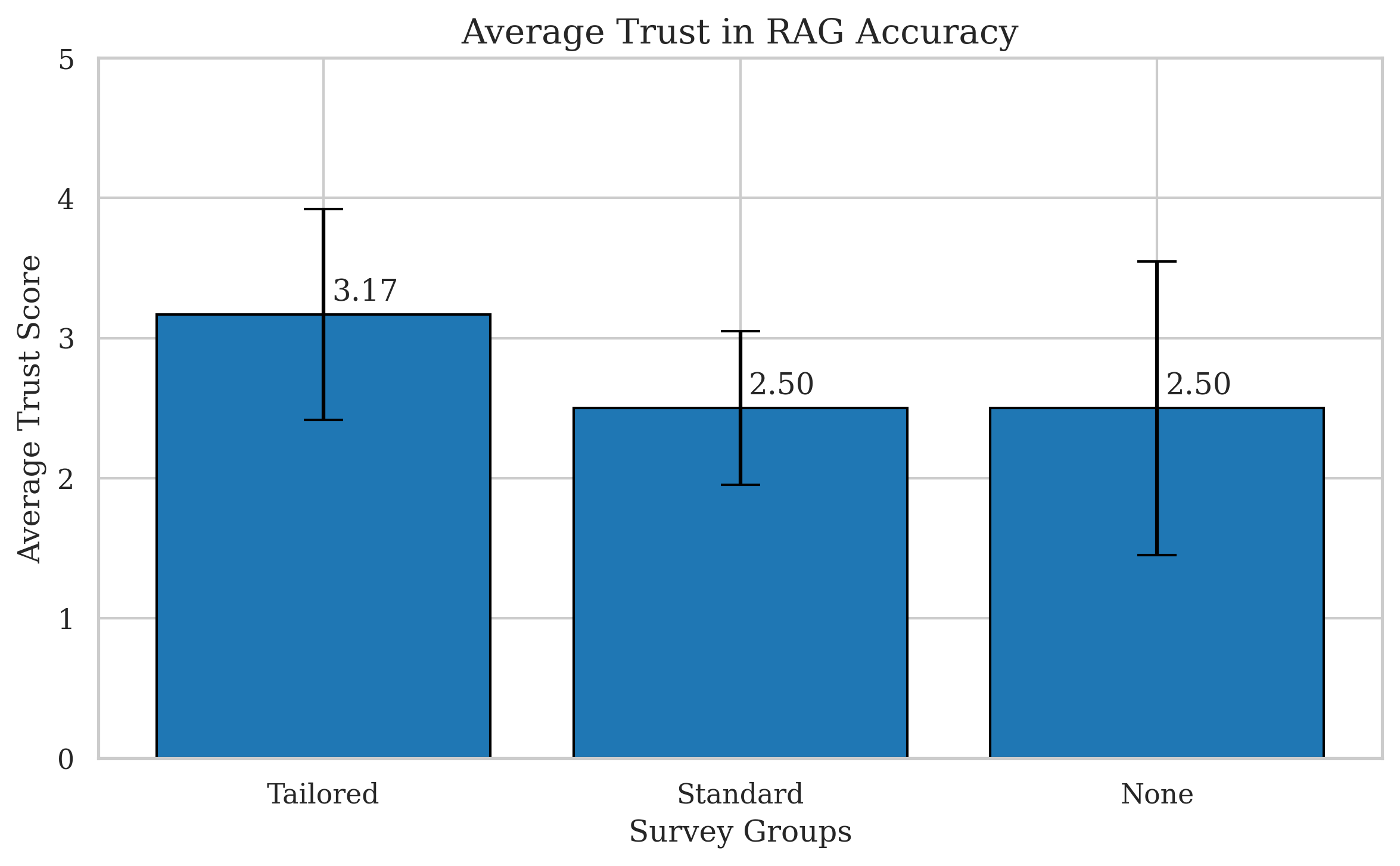}
  \caption{The average trust (in scale of 1-5; 1 is the lowest, 5 is the highest) to the system reported by participants of the pilot study.}
  \label{fig:trust}
\end{figure}

\begin{figure}[h]
  \centering
  \includegraphics[width=0.8\textwidth]{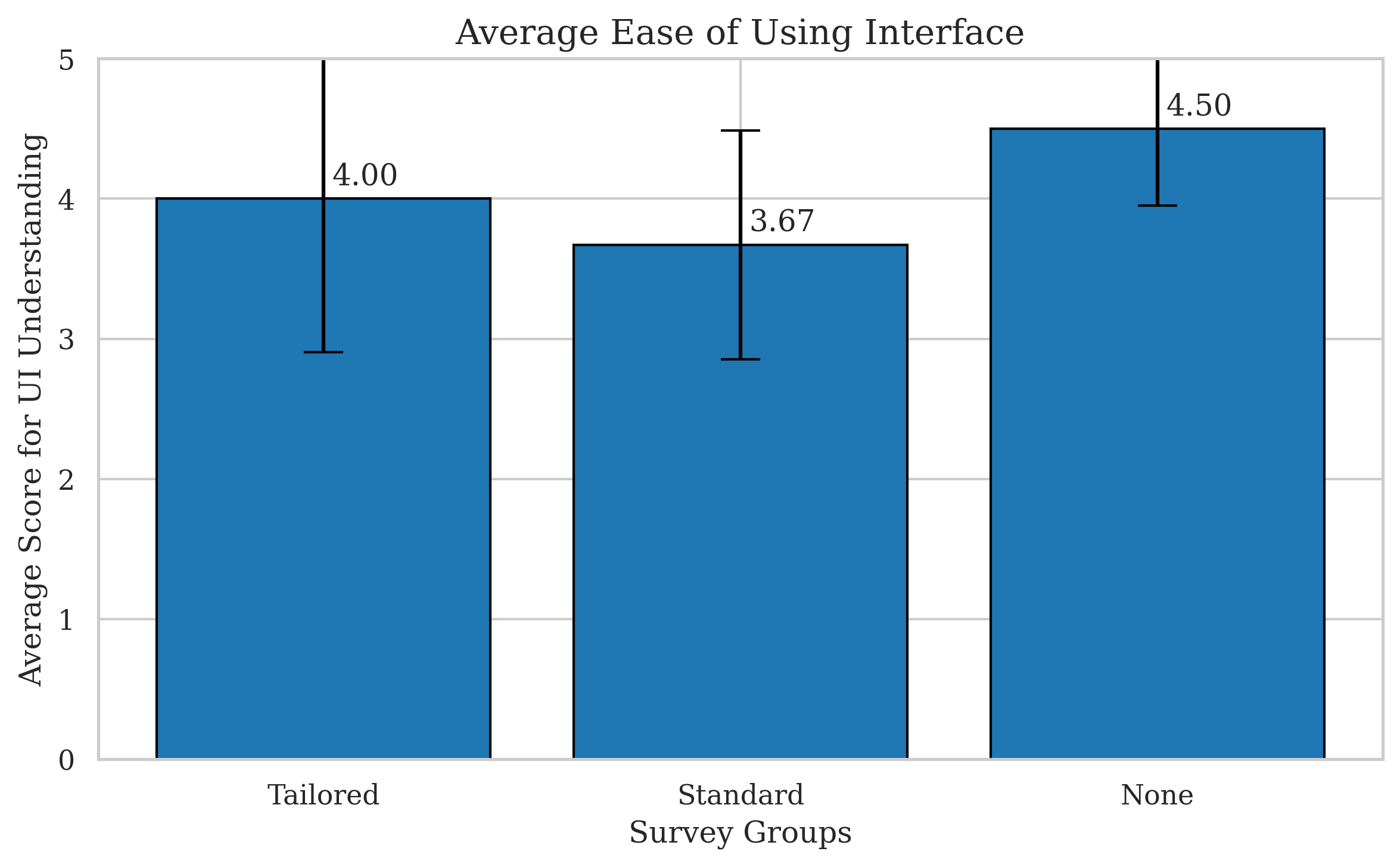}
  \caption{The average ease (in scale of 1-5; 1 is the lowest, 5 is the highest) to the system reported by participants of the pilot study.}
  \label{fig:ease}
\end{figure}

\textbf{Trust and User Experience.}
The evaluation of interface usability (Figure \ref{fig:ease}, scored from 1 (not easy to use) to 5 (very easy to use), shows that participants in the no warning group found the interface easiest to use, followed by the tailored warning group, and finally, the standard warning group. This might suggest further improvements can be made in the way warning messages were conveyed to the user; that said, it is certainly interesting how the tailored survey group had a easier time utilizing the interface than the standard. A possibility for this result may lie in that more useful/specific warning messages allows for the user to feel as if they understood the output better, thus feeling more at ease. Perhaps more importantly, Figure \ref{fig:trust} expresses how tailored users tend to have more trust in the model than their other two counterparts, with a difference of 0.67 (out of a 5-point Likert scale).

\section{Discussion}
The nuanced relationship between user trust and warning systems reveals a complex psychological dynamic in human-AI interaction \cite{zerhoudi2024} \cite{wang2024farsight}. Our findings suggest that transparency is not merely about presenting warnings, but about crafting them in a way that empowers users rather than intimidates them. The statistically significant difference in trust levels (0.67 on a 5-point scale) between the tailored warning group and other groups underscores the critical role of contextually relevant information \cite{nahar2024fakes}.

While this study focuses on educational contexts, the cognitive impact of tailored warnings extends to other AI-augmented reasoning applications, including healthcare, law, and finance. In these high-stakes fields, where users rely on AI for critical decision-making, effective warning systems play a crucial role in ensuring that AI serves as a reasoning aid rather than an unquestioned authority. Increasingly, users seek systems that not only provide information but also offer meaningful insights into its reliability \cite{byun2024} \cite{tan2024retrieval}. The tailored warning approach represents a shift from passive information consumption to active engagement with AI-generated content. By delivering specific, actionable context about potential hallucinations, these warnings empower users to critically evaluate information rather than passively accept it \cite{sudhi2024rag}.

Moreover, the psychological impact of these warnings cannot be overstated. Traditional approaches often create a binary perception of AI systems as either entirely trustworthy or completely unreliable \cite{wu2024clasheval}. Our research demonstrates a more nuanced approach: users can be made aware of potential limitations while maintaining a constructive relationship with the technology. This approach aligns with emerging research on responsible AI development, which emphasizes transparency and user empowerment \cite{kumar2024decoding}. During the qualitative portion of the study, one participant questions, "Why can't you just provide us the right answer if you know how to warn us?" Another participant mentions, "The warning messages confuse me. I just want the correct answer." These reactions raise the question: under what context is pushing users to "think critically" with warning messages desirable? If users are naturally prone to clear-cut answers, what would be the best way to emphasize the biases and problems of AI system outputs? Future studies in this space also need to draw more from psychology and cognitive science research to inform the best system design that effectively fosters critical thinking.






\bibliographystyle{ACM-Reference-Format}
\bibliography{sample-base}

\end{document}